# Specialty Fibers for Terahertz Generation and Transmission: A Review (*Invited*)

Ajanta Barh, B. P. Pal, *Senior Member*, *IEEE*, G. P. Agrawal, *Fellow, IEEE*, R. K. Varshney, and B. M. A. Rahman, *Senior* Member, *IEEE*

*Abstract*—Terahertz (THz) frequency range, lying between the optical and microwave range covers a significant portion of the electro-magnetic spectrum. Though its initial usage started in the 1960s, active research in the THz field started only in the 1990s by researchers from both optics and microwaves disciplines. The use of optical fibers for THz application has attracted considerable attention in recent years. In this article, we review the progress and current status of optical fiber-based techniques for THz generation and transmission. The first part of this review focuses on THz sources. After a review on various types of THz sources, we discuss how specialty optical fibers can be used for THz generation. The second part of this review focuses on the guided wave propagation of THz waves for their transmission. After discussing various wave guiding schemes, we consider new fiber designs for THz transmission.

*Index Terms*—Optical waveguides, plastic optical fibers, THz generation, THz transmission

## I. INTRODUCTION

THE TERAHERTZ (THz) frequency/wavelength window of the electro-magnetic spectrum lies between the infrared band and the microwave band (see Fig. 1), and ranges in frequency from 0.1 to 10 THz (equivalently wavelength ranges from 3 mm – 30 μm) [1], [2]. Radiation from any object with temperature > 10 K contains THz wavelengths and almost 98% of cosmic background radiations since Big-Bang event corresponds to THz and far-infrared frequencies [3], [4]. Initially THz radiation was mainly used for *passive* applications, where THz waves were detected to study the chemistry of cold planetary atmosphere and interstellar medium [4]-[6]. It even had different names—sub-millimeter or extreme far-infrared [6], [7]. Over the last 20 years, THz technology has experienced a significant growth, almost in an exponential manner [8] owing to its potential applications. This frequency range can transfer huge data files via wireless route [9], [10] and can significantly increase the communication data rate over existing microwave technology [9]-[12]. The THz radiation scatters less than optical waves, is non-ionizing, and can yield high resolution images [1], [3], [13]-[15]. THz waves can deeply penetrate through cloths, ceramics, walls, woods, paper, dry air, polymers etc., but they are absorbed or reflected by metal, water vapor, dust, cloud, and sufficiently dense objects [3], [16]-[18]. Moreover, vibrational or rotational transitions of a wide range of molecular clusters, and electronic transitions of many nano-composites (chemical and biological substances) exhibit strong resonances at THz frequencies, and those absorption bands can serve as their *chemical fingerprints* [19]-[24]. Additionally, THz waves are also not harmful to human health [25]-[27]. All these excellent qualities of THz radiation make it suitable for imaging of hidden objects, like explosives, metallic weapons etc. [28]-[30]. It is also useful to monitor the buried defects in IC packages, layer of paints, tiles, or space crafts etc. [31]-[34]. Most importantly, this technology has already begun to make deep inroads in non-invasive medical diagnostics, such as detection of skin cancer, tooth decay, and identification of human tissues based on different refractive indices and linear absorption coefficients at THz frequencies [1], [35]-[38].

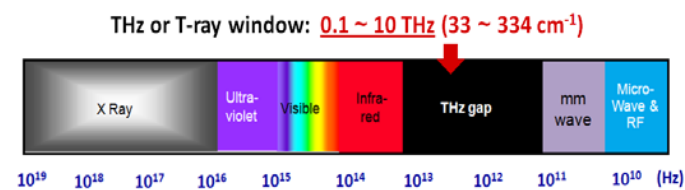

Fig. 1. Schematic diagram of electro-magnetic spectrum, indicating THz gap between far-infrared and mm wave region. After, Ref. [31].

THz waves suffer high metallic losses in conventional electronic circuitry, which operates below 100 GHz. It is well established that for typical radio frequency devices, like electronic circuits, the efficiency as well as the average output power tends to fall inversely with second to fourth power of the operating frequency [2], [35], [39]. Moreover, this high loss and low-efficiency along with the smaller size of electronic devices lead to another problem, the extreme heating effect [39], [40]. An expensive cryogenic/cooling system is eventually required for dealing with this heating

Manuscript received June, 2015. Partial funding by Trilateral UKIERI project entitled "Design and analysis of optical microstructured fiber-based THz waves for transmission and applications" is gratefully acknowledged. A.B. acknowledges the support of CSIR.

Authors Ajanta Barh and R. K. Varshney are with the Department of Physics, Indian Institute of Technology Delhi, Hauz Khas, New Delhi 110016 India (e-mail: ajanta.barh@gmail.com, ravi@physics.iitd.ac.in, respectively).
Author B. P. Pal is with the School of Natural Sciences, Mahindra Ecole Centrale, Hyderabad 500043 India (e-mail: bishnupal@gmail.com).
Author G. P. Agrawal is with Institute of Optics, University of Rochester, Rochester, NY 14627 USA (e-mail: govind.agrawal@rochester.edu).
Author B. M. A. Rahman is with Department of Electrical and Electronic Engineering, City University London, London EC1V 0HB UK (email: b.m.a.rahman@city.ac.uk).

issue [40]. Though use of THz waves has significant benefits, THz generation is limited by lack of suitable narrow band-gap semiconductors [41], [42]. The band-gaps corresponding to lower THz frequencies (< 10 THz) become comparable to the lattice vibration energy, and hence they are associated with huge thermal noise [41]. The conventional dielectric waveguides such as silica glass fibers are also not useful for guiding THz wave owing to their high transmission losses [43]. Moreover, suitable electro-optic devices are also not available till date. However, the researchers are trying hard on various technology options to fill this THz-gap!

Recent advances in THz technology are motivated by a rapid growth of THz-based *active* applications mentioned above earlier. The primary challenges are to design and develop appropriate THz sources, low-loss transmission medium, efficient detectors, and other control components [41], [43]. Among these, the most critical one is an efficient, high power, portable and stable THz source [39], [41], [44]. A low-loss transmission medium is also desirable to reduce the overall loss for many applications [43]. The detection of THz wave was performed first in 1890s by using the bolometer [45], [46]. Since then, several proposals have been made [41], [47]-[52], though they were quite expensive and required special cooling system. However, the recent development in THz time-domain spectroscopy has boosted the bolometer detection technique significantly [22], [53], [54].

The researchers from both optical and electrical disciplines are striving hard to achieve more and more efficient THz technology by continuously improving the existing devices as well as developing new ones [39], [55]. In this review article, we focus on the *optical* side of THz technology. Since, it itself forms a vast domain, we concentrate mostly on *optical waveguide* (OWG) based THz sources and low-loss transmission systems. Research on OWG-based applications began around 2000 [56]-[58]. Among various OWG platforms, *optical fibers* are most attractive for day-to-day applications, which include high power lasers, long distance transmission, medical endoscopy, sensing etc. [59]-[61]. Moreover, they can be packaged in a reasonably compact form [62]. Several fiber-based innovative proposals have already been made, especially for loss-less THz guidance.

This review article is organized as follows. In section II, we first survey the schemes for generating THz wave via optical means, and then focus on optical fiber-based proposals. In section III, we investigate dielectric fiber based designs for THz transmission. Last section provides future prospects.

## II. Terahertz Sources

As mentioned above, plenty of THz radiation is available in nature, including cosmic or interstellar background radiation, black-body radiation of any object with temperature > 10K, ordinary mercury lamp, heated rod of carborundum etc. [3], [8], [63], [64]. However, these are hardly suitable for *active* applications mentioned earlier. The appropriate source design for the THz band is certainly the most challenging task. To date, several techniques have been proposed, and depending on their working principle they can be divided in three broad categories: (1) particle accelerator based sources, (2) microwave/electronics based techniques, and (3) optics based techniques. Till now, highest average power has been obtained from the 1$^{st}$ category, which can also provide sources with a continuously tunable wavelength. Unfortunately, these are not cost effective owing to their very large size and complex set up. Few examples are backward wave oscillators, travelling wave tubes, extended interaction klystron, gyrotrons, free electron lasers, synchrotron etc. [65]-[68]. In the second category, electronics based THz sources are undoubtedly the more compact ones and can generate CW power with a narrow line width at room temperature. Though they work fine in the lower frequency end, their efficiency as well as output power drops rapidly as THz frequency increases. In that regime, their metallic structures suffer high losses too [2]. Few examples of electronics-based THz sources are Gunn diodes, high frequency transistors, etc. [69]-[71]. In the third category, *optics* based sources yield the highest average power at the high end of the THz frequency range. As the operating frequency decreases, thermal noise becomes an unavoidable issue, and consequently, cryogenic cooling systems become necessary. Though a wide range of optical based THz sources are now available, they still lack in efficiency. Another challenge is to find suitable dielectric material as a host for THz operation. Recently, superconductor based THz sources are proposed by exploiting Josephson junctions [72]. In the following we will focus mainly on fiber-based THz sources.

### A. Review of Optical THz Sources

Depending on the state of operating medium, optical THz sources can be divided in two categories: gaseous and solid-state THz sources. Among these, gas lasers are the oldest ones [73], [74], whereas quantum cascade lasers (QCL) are the newest one [75].

Gas lasers emit THz waves within a wide spectral range as no phonon resonances occur in a gaseous medium. They also prevent any reflection of both the optical pump and generated THz waves. Several organic (e.g. methyl chloride, methanol, vinyl chloride etc.) and inorganic gases (e.g. air, nitrogen, helium, argon, krypton etc.) at low pressure have been used with an intense optical pump (usually $CO_2$ laser) to generate broad-band THz waves [73], [76], [77]. The resulting THz sources are coherent and can yield sufficiently high average power. One noticeable drawback is in their bulky set up; however, their efficiency is still low though steadily improving [77]-[79].

Solid state optical THz sources can be categorized in two classes, one is a *laser* source and the other is a *laser pumped* source. The THz-laser sources, such as semiconductor lasers and QCLs, are frequency-tunable and quite compact in size. Such sources include germanium (Ge) or silicon (Si) based laser [80], strained *p*-Ge laser [81], and electrically pumped photonic-crystal laser [82]. The QCLs are primarily composed of hetero-structured materials (Si/SiGe, GaAs/AlGaAs, InGaAs/InAlAs, GaSb/AlGaSb) with multi-quantum well structures [83]-[89]. They have become quite popular for high THz frequency operation, and their output power varies from

1 µW to 10's of mW level depending on its operating frequency. Unfortunately, they too require a costly cryogenic cooling system. Till date, lowest THz frequency realized with a QCL is 1.2 THz with CW output power of 0.12 mW [90] and the highest operating temperature is 199.5 K [91]. Research is still going on to improve both these issues [92], [93]. In the second category, a wide variety of laser pumped THz sources had been proposed, such as a photoconductive or dipole antenna [94]-[98], a photomixer [99]-[102], a nonlinear crystal or polymer based difference-frequency generator [44], [103]-[106], an optical rectification based source [107]-[110], a THz parametric oscillator [111]-[114], or an enhanced surface emitter in a magnetic field [115], [116]. Though research is still continuing in these fields, the maximum efficiency is still below 0.05%. However, in terms of compactness, tunability and CW operation, such THz sources are quite promising [117]. Other approaches include magnetic component of light induced charge separation in dielectric [118], optical fiber based application etc. In the next section, we will study this optical *fiber* based THz generation in detail.

### B. Optical Fiber Based THz Generation

THz generation using optical fibers is attractive owing to their versatile characteristics. Most of the proposals are based on conventional fused silica optical fibers [119]-[121], but recently polymer fiber based THz generation has been also proposed [122]. In all cases, nonlinear effects are exploited to generate the THz waves with an optical source acting as the input pump. Recently an interesting idea for THz generation has been reported based on photo-Dember effect [123]. In this approach, a thin film of InAs semiconductor material is coated on the carefully polished end facet of the fiber. The fiber guides pump light to this film and THz radiation is actually generates inside it.

The nonlinearity in optical fibers arises from the third order susceptibility [$\chi^{(3)}$] [124], which is commonly exploited to generate THz waves [124]-[126]. Among the several nonlinear physical processes, four wave mixing (FWM) is the most relevant one for new frequency generation, provided appropriate phase matching condition can be satisfied [124], [127]. In the FWM process, two pump photons ($\omega_p$) of the same (degenerate FWM) or different frequencies (non-degenerate FWM) are converted to a signal photon of lower frequency ($\omega_s < \omega_p$) and an idler photon of higher frequency ($\omega_i > \omega_p$), as shown in Fig. 2: the subscripts p, s and i stand for pump, signal and idler, respectively. Along with energy conservation ($\omega_{p1} + \omega_{p2} = \omega_s + \omega_i$), momentum conservation must also be simultaneously satisfied for fulfilling the required *phase matching condition*. By launching a weak idler as a seed at the input along with the pump(s), the FWM efficiency can be improved considerably, as it stimulates the FWM process.

Depending on the momentum conservation, the THz wave can be generated either parallel to the input optical waves [*collinear*, cf. Fig. 3(a)] or perpendicular (or any other angle) to them [*non-collinear*, cf. Fig. 3(b)]. In Fig. 3, $\beta_j$ is the propagation constant of the $j^{th}$ mode (j = p, s, i). It should be noted that, along with the generation of a THz wave, input seed (idler) will also get amplified [126]. Under the collinear phase matching condition, the THz wave can be collected at the end facet of the optical fiber [119], [122], whereas for non-collinear phase matching, THz is emitted sideways from the cladding of the fiber [119], [120]. The latter is attractive for bio-medical applications.

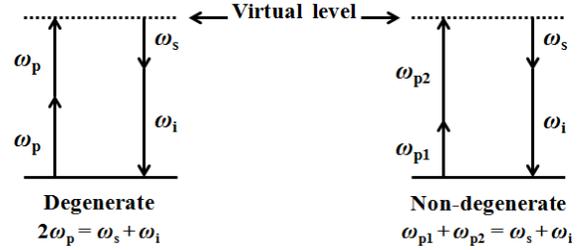

Fig. 2. Schematic energy level diagram of the degenerate and non-degenerate FWM processes.

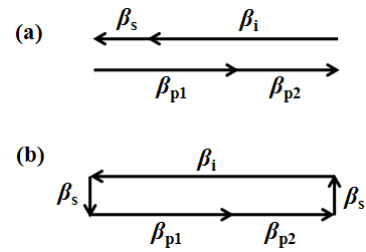

Fig. 3. Schematic of (a) collinear and (b) non-collinear phase matching conditions.

The frequency shift ($\Omega_s = \omega_p - \omega_s$) strongly depends on both the group velocity dispersion (GVD) and nonlinear parameters of the fiber at the pump wavelength ($\lambda_p$), which should ideally lie close to the designed fiber's zero dispersion wavelength [127]. Dispersion parameters are obtained by expanding $\beta(\omega)$ of the fiber in a Taylor series around the pump frequency and are given by $\beta_n = d^n\beta/d\omega^n$. By considering dispersion terms up to $5^{th}$ order, $\Omega_s$ can be approximated as [124], [126]

$$\Omega_s = \sqrt{\frac{6|\beta_2|}{\beta_4}\left(1 \pm \sqrt{1 - \frac{2\beta_4\gamma P_0}{3\beta_2^2}}\right)}, \qquad (1)$$

where, $\gamma = (2\pi n_2)/(\lambda A_{eff})$ is the nonlinear parameter and $P_0$ is the input pump power (degenerate case). The parameter $n_2$ is known as nonlinear index coefficient and is related to $\chi^{(3)}$. The FWM efficiency strongly depends on the following parameters:

- Residual phase mismatch
- Modal overlap among participating waves
- Effective nonlinearity of the fiber
- Fiber's losses
- Similarity in polarization states
- Single mode operation at pump and signal frequency

Conventional optical fibers (fused silica) exhibit relatively low losses up to ~ 2 µm [128], but become highly lossy at THz wavelengths, which make them unsuitable for THz propagation. One sure way to improve the output power of THz radiation generated inside a fiber is to let it radiate out from the fiber core. Such fiber can be designed by employing the *non-collinear* phase-matching condition. This was first proposed by *K. Suizu et al.* in 2007 [119], when they theoretically investigated *surface emitted* THz waves from a silica optical fiber. As seen in Fig. 3, this approach requires the pump and idler to propagate in counter directions. The phase matching is achieved by considering the *uncertainty* in momentum of THz wave because of a small size of the fiber core (radius is few μm), which is a small fraction of the THz wavelength > 100 μm). The phase matching wavelengths and the output THz power were calculated by solving nonlinear equations analytically by assuming negligible pump depletion and fiber losses. Under such conditions, the output THz power ($P_{THz}$) can be expressed as [119]

$$P_{THz} = \frac{16 n_{THz}^5 n_2^2}{9 n_o^3 \lambda_{THz}^3 w} \left( \mu_0 / \varepsilon_0 \right) L P_1 P_2 P_i \qquad (2)$$

where, $n_{THz}$, $n_o$, $\lambda_{THz}$, and $w$ are the modal index of THz wave, modal index of optical wave(assumed same for all optical waves), wavelength of the THz wave, and the mode field radius ($1/e^2$) of optical wave, respectively. The parameters $\mu_0$ and $\varepsilon_0$ are the free-space permeability and permittivity, respectively. $L$ is the total fiber length, and, $P_1$, $P_2$, and $P_i$ are the input powers of the two pumps and the idler, respectively. The reported results in [119] reveal that by using two optical sources, one at a wavelength of ~ 800 nm ($\omega_i$) and other at 1.55 μm ($\omega_{p1} = \omega_{p2}$), a THz wave can be generated at the frequency of ~ 2.6 THz. By tuning the pump wavelength from 1.48 to 1.62 μm, the THz frequency can also be correspondingly tuned from 2.52 to 2.64 THz. In another paper [120], similar approach has been exploited to generate surface-emitted THz waves, where a more detail derivation of output THz power has been presented. This radiated THz power can be collected by wrapping the fiber around a bobbin (plate/cylindrical) and then by focusing them collectively through suitable lenses (cf. Fig. 4). Thus, the total power can be enhanced by using long fiber length. However, the efficiency of this design is still quite low (~ $10^{-6}$).

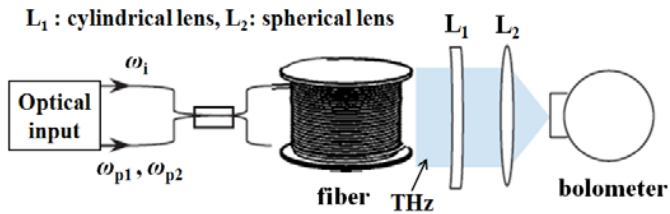

Fig. 4. Schematic diagram of surface-emitted THz wave from optical fiber. After, Ref. [120].

Collinear phase matching can be achieved through several routes, such as periodic poling [129], a Bragg-grating geometry [121], or a suitable nonlinear phase shift [127]. THz generation using those techniques has already been studied in both silica fibers [119], [121] and plastic fibers [122].

In case of silica fibers, the main drawbacks of *collinear* phase matching are high material loss and a small core size (compare to the THz wavelength). A small core increases the confinement loss or may not support any THz-frequency mode at all! Interestingly, the THz wavelength is comparable to the size of a fiber's cladding. One solution would be to guide the THz wave through fiber's cladding. Theoretical results have been reported for such a configuration [119]. It was found that a tunable THz source in the frequency ranges of 6.93–7.33 THz and 1.07–1.12 THz is realizable by tuning the wavelength of the input pump in the range of 1.48–1.62 μm. Fiber Bragg grating (FBG) based *collinear* phase matching configuration has also recently been reported in [121], where a FBG is used to compensate the phase mismatch among the interacting waves. In this configuration, the efficiency of nonlinear process can be improved by applying an external static electric field such that the direction of applied field is parallel to the direction of polarization of input beams. This external electric field induces a $2^{nd}$ order nonlinear effect in the fiber (through Pockel's effect). The strength of this nonlinearity can be enhanced by increasing the external voltage. By launching two optical sources (1.555 and 1.588 μm), a THz wave (~ 4 THz) could be generated through difference frequency generation (DFG) [44], [121], [124], [130]. Here, the phase matching condition is satisfied only for DFG, and other nonlinear effects such as, second harmonic, third harmonic, and sum frequency generation can be neglected. Tuning range of the generated THz wave depends on the phase matching band width, which in turn depends on the dispersion properties of the FBG as well as on the input waves. Effects of fiber losses have been reduced by considering a small fiber length (~ 3 cm). As the input optical waves interact under DFG, they generate THz wave, which radiates out from the end facet of the fiber and collected by a lens. However, the efficiency is still quite low (~ $10^{-4}$) for this configuration [121].

Recently, we have proposed a design of a high power THz source based on a specialty *plastic fiber*, where degenerate FWM with *collinear* phase matching has been exploited to generate THz waves [122]. Our primary aim was to improve the efficiency by launching pump light from a commercially available mid-infrared source into a dispersion engineered plastic fiber. Microstructured optical fibers (MOF), with micron scale (order of operating wavelength) features across their cross-section, are often used for engineering dispersion and nonlinearity [60], [124]-[127], [131]-[134]. On the other hand, *plastic* material, such as Teflon, PC, HDPE, PE, PMMA, COC etc., are highly transparent and possess flat material dispersion in the THz range [17], [135]. Moreover, they can be drawn in a fiber form, and researchers have already reported their successful fabrication as MOF structures [43], [136]-[138]. From the literature survey it is also evident that several polymer materials exhibit high optical Kerr nonlinearity when prepared with proper processing and doping ($n_2$ ~ $10^{-18}$ – $10^{-17}$ m$^2$/W) [139], [140]. Combining all these excellent properties, we considered a Teflon based MOF for generating the THz wave. Since high power $CO_2$ laser emitted at ~ 10.6 μm are available commercially, we use this laser for pumping the degenerate FWM process. Our target is to enhance the FWM efficiency by increasing the *modal overlap*. Since wavelengths of three waves involved in the FWM process are very different, it is a challenge to realize single-mode operation with high modal overlap. We use a *double clad* fiber geometry shown in Fig. 5(a) to balance all the criteria. In this Teflon based *microstructured core double clad plastic fiber* (MC-DCPF), the central core is formed by using 4 rings of hexagonally arranged higher index rods ($n_r$, Teflon-type-1 shown as white dots in Fig. 5(a)) embedded in a lower index background ($n_b$, Teflon-type-2, dark blue in color), which also forms the uniform cladding (1st cladding) for optical waves. These two regions (composed of $n_r$ and $n_b$) together form the microstructured-core based large mode area fiber geometry for the optical modes (similar to that used in [141]), which essentially increases their mode field diameter. Refractive indices $n_r$, $n_b$ and structural parameters, such as, radius of rods ($r$), spacing between rods ($\Lambda$) and the radius up to 1st cladding ($R_1$) were optimized to get desired dispersion profile, nonlinearity, effective single mode operation, low-loss etc. Interestingly, this whole region (up to $R_1$ in Fig. 5(a)) acts as the waveguide core for the THz wave ($\lambda \sim 100$ μm) and its cladding is formed by adding an extra layer of Teflon (Teflon-type-3, refractive index $n_{cl}$, light blue in color) around it. The outer boundary of this cladding (radius $R_2$) is chosen to get negligible confinement loss at THz wavelengths. Using this configuration, we have shown that a high-power pump (at 10.6 μm with 1 kW power) generates a THz wave at a wavelength near 100 μm (i.e., frequency ~ 3 THz) with an idler wavelength at ~ 5.6 μm. This idler corresponds to that of a CO laser (available commercially), which can be used as a seed for the FWM process. The important thing is that our designed MC-DCPF geometry provides a very high *modal overlap* among all the three participating waves (pump, seed and THz), which is the key factor to increase the power transfer efficiency from pump to THz wave. The 1D intensity profiles of the fundamental pump and THz modes are shown is Fig. 5(b). They were obtained by calculating the Pointing vector of the corresponding modes. The output power of THz waves is calculated by rigorously solving the FWM coupled equations assuming CW operation for all the waves [122]. Both fiber loss as well as pump depletion have been incorporated in this study. We have shown that, if we launch a pump with 1 kW power along with the seed power of ~ 20 W power, a THz wave with 30 W power is generated at the end facet of a 65 m long MC-DCPF with the power conversion efficiency of 3%, which is more than 100 times higher than any other proposal to date on THz sources. The output phase matching band-width of the generated THz wave is plotted in Fig. 6. The 65-m length of our plastic fiber is relatively long. However, the fiber can be easily spooled around a cylindrical bobbin to make a compact device, and for a bobbin radius ≥ 10 cm, the bend-loss is negligible.

Table I provides a summary of fiber-based schemes discussed in this section for generating THz waves. It is clear from this table that the efficiency, band-width, and the output THz power are growing gradually. This trend may continue by exploring newer ideas and different fiber structures, made of suitable materials. However, we may stress that these proposals are still at a theoretical stage. The real challenge is to fabricate and use them for real applications.

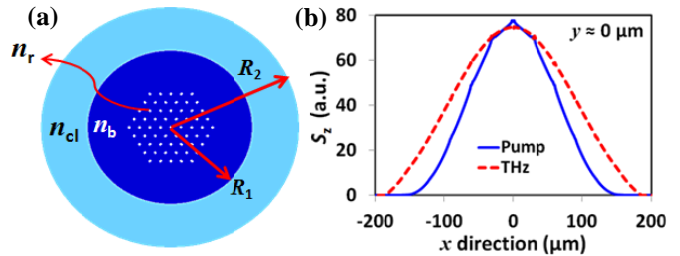

Fig. 5. (a) The schematic cross-section of the proposed MC-DCPF. The central white rods are composed of Teflon-1 (refractive index $n_r$), the dark blue background is made of Teflon-2 ($n_b$) and the light blue outer region (cladding for THz wave) is made of Teflon-3 ($n_{cl}$). (b) Mode intensity profiles for the pump and THz waves along *x*-direction (*y* = 0). After, Ref. [122].

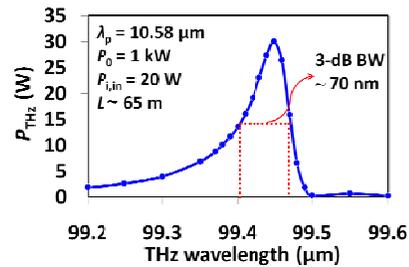

Fig. 6. The spectrum of generated THz wave at the end of fiber output facet. 3-dB phase matching band-width is ~ 70 nm. After, Ref. [122].

TABLE I
THZ GENERATION IN FIBER STRUCTURE

| Generation scheme | Input pump (Optical source) | | THz output | | | Ref. year |
|---|---|---|---|---|---|---|
| | Power | λ (μm) | Type | f (THz) | efficiency | |
| FWM in silica fiber | 1 kW (each) | 1.48-1.62, 0.8 | Pulse/CW | 2.52-2.64, 6.93-7.33, 1.07-1.12 | ~ $10^{-5}$ | [119] 2007 |
| -do- | -do- | -do- | Quasi CW | 3.15-3.25 | ~ $10^{-6}$ | [120] 2011 |
| DFG in silica fiber | 5 - 35 W | 1.555 & 1.588 | CW | 2 – 30 (peak ~ 4) | ~ $10^{-4}$ | [121] 2012 |
| Photo-Dember effect | 0 ~ 17 mW (ave.) | 0.8 | pulse | 0.1 ~ 2 | - | [123] 2010 |
| FWM in Teflon MOF | 1 kW, 20 W | 10.58, 5.59 | CW | Centered ~ 3 | ~ $10^{-2}$ | [122] 2015 |

## III. GUIDED TRANSMISSION OF THZ

Transferring information via electromagnetic fields requires low-loss and distortion-free wave guiding schemes. Metal wires have been used for centuries for electronic transmission [142], whereas, dielectric waveguides have been employed for decades for optical communication [143]. In the case of THz waves, researchers are still searching for suitable wave guiding scheme(s).

### A. A Brief History of THz Waveguides

Dry air is perhaps considered to be the best medium for THz transmission in terms of both low-loss and low-dispersion. However, the presence of water vapor, dust, cloud etc. limits its transmission and exhibits several strong absorption lines over the THz band [18]. Hence, although most of the present THz technology relies on *free space* transmission and manipulation, the propagation distance is limited by atmospheric losses, which may vary with weather, climate change, geographical location, altitude etc. For this reason, appropriate THz waveguides (WGs) are important for a number of applications, such as remote sensing, long range communication, collimated and diffraction limited beam guiding. Moreover, a THz waveguide can be used as an imaging probe (e.g. in medical endoscopy), or its strong confinement property could be explored to enhance the light-matter interaction. The primary barrier for this technology is the lack of a suitable *material* for the WG structure. Several proposals have already been made in this direction, based on, *metallic* WGs, *dielectric* WGs and in some cases metal-dielectric *hybrid* WGs as discussed below.

Early THz WGs were composed of *planar* structures where coplanar transmission lines are quite common for microwaves [144], [145]. Unfortunately, they suffer high propagation losses (~ 20 cm$^{-1}$) due to the combined effects of Ohmic loss, absorption loss inside a dielectric substrate, and radiation loss. For this reason, *non-planar* WG structures became an attracting option. Both metallic (cf. Fig. 7) as well as dielectric based non-planar structures have been proposed in literature. In their early forms *non-planar metallic* WGs consisted of a circular structure [cf. Fig. 7(a)]. In these, THz modes (both TE and TM) could be confined tightly inside their metallic walls. Such WGs are nearly free from radiation and dielectric losses but still suffer from high metallic losses (~ 0.7 cm$^{-1}$ at 1 THz) and high dispersion near the cut-off frequency. A detailed description of such circular WGs and rectangular WGs is given in [146]. Parallel plate WGs [cf. Fig. 7(b)] support a TEM mode, which has no cut-off and hence doesn't suffer from dispersion. These WGs exhibit lower loss than circular and rectangular WGs [147]. In 2004, the lowest loss (0.03 cm$^{-1}$) THz transmission through a bare metal wire [cf. Fig. 7(c)] was demonstrated [148] and applied for endoscope application. In a properly designed wire, the THz mode attaches loosely to it (acts as a rail) and spreads in the surrounding medium. This configuration reduces the metallic loss greatly, but the mode becomes less confined. Coupling of a free-space THz mode to this metal wire is also difficult. A two-wire based THz wave guiding scheme [cf. Fig. 7(d)] was proposed in 2010 [149], which supports TEM as well as TE and TM modes; dispersion becomes negligible for the TEM mode. Moreover, the coupling loss reduces significantly, as the TEM mode resembles well with the THz mode generated from a dipole antenna. This geometry confines the THz wave between two wires; however this confinement decreases for a larger radius of metallic wires and for larger separation between them. Another type of geometry, called slit WG [cf. Fig. 7(e)], provides higher mode confinement (0.1 ~ 1 THz) with nearly no dispersion [150]. Incidentally, such high confinement also leads to high metallic losses. In conclusion, though several advances have been made towards designing THz-metallic WGs, there exist a strong trade-off between mode confinement and metallic loss. Moreover, most of the proposals are only suitable for short distance transmission (mm length scale).

For integrated circuit applications *hybrid plasmonic* WG structures have become quite popular. A plasmonic WG can confine electromagnetic wave very tightly (sub wavelength scale) in the form of a surface plasmon mode at the interface of a *metal* and a *dielectric* medium [151]. This kind of WG has also been investigated for THz waves [152]-[154]. Both the in-plane as well as out-of-plane confinement of THz mode is possible, and its dispersive properties can be tuned arbitrarily by engineering the interface design.

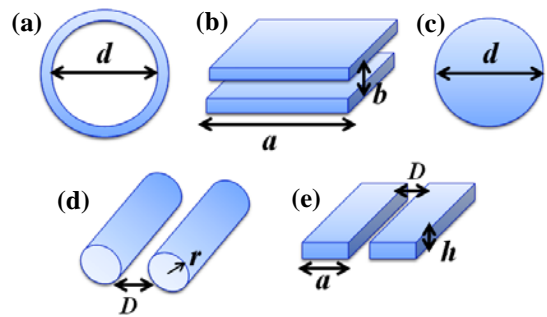

Fig. 7. Schematic diagram of non-planar metallic waveguides. (a) Circular, (b) parallel plate, (c) bare metal wire, (d) two-wire, and (e) slit waveguide.

A *dielectric* WG mainly suffers from frequency dependent material absorption loss. High-resistivity (HR) silicon possess extremely low loss and negligible dispersion up to ~ 3 THz [155], however, it is only useful for chip scale applications. As an example, researchers have theoretically designed HR-silicon based split WG structure (similar to Fig. 7(e), except that the WG is made of dielectric) for tight confinement of THz waves [156]. Alternatively, as discussed earlier, a number of plastic materials, such as Teflon, HDPE, PE, PMMA, COC etc., are highly transparent and possess flat material dispersion over the THz range [17], [135], [157]. Moreover, they can be drawn into a fiber form. Several proposals have already been made towards designing plastic based THz WGs. However, in the following we would focus on *fiber* based THz guidance.

B. *Optical Fibers for THz Transmission*

Electromagnetic waves can be confined to dielectric WGs via index guiding (IG), photonic band-gap (PBG), or an anti-resonance guiding mechanism [59]. All of them have already been explored for making *fiber* structures for THz transmission.

The core of an *IG-fiber* has higher refractive index than the surrounding cladding region, and light is guided in the core via total internal reflection at the core cladding interface. Three kinds of cross-sectional geometries have been proposed for IG-fibers in literature. One is a conventional step-index fiber, where the core as well as the cladding regions is composed of uniform materials [cf. Fig. 8(a)]. The second type is a solid core microstructured fiber [cf. Fig. 8(b)], where a uniform core is surrounded by a cladding, consisting of micron scale (order of operating wavelength) periodic or non-periodic refractive index features that run along the fiber length. The third type is a porous/microstructured-core fiber, where the core is composed of micron-scale refractive index features, surrounded by a uniform cladding [cf. Fig. 8(c)]. Note that we have already discussed a THz source exploiting the third kind of fiber geometry [cf. Fig. 5(a)].

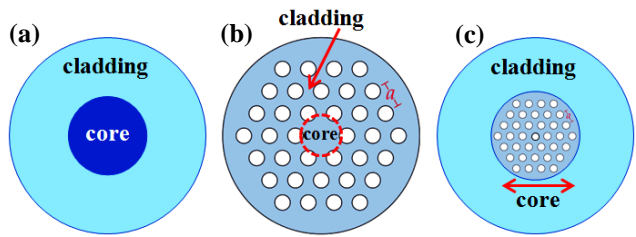

Fig. 8. Schematic cross-sectional diagram of different TIR-guided fiber structure. (a) Step-index fiber, (b) solid core microstructured fiber, and (c) porous-core fiber. The white circles in (b) and (c) are air holes, which can also be filled by other lower index (compare to core index) material(s).

The first dielectric WG for THz guiding was proposed in the year 2000, where a plastic (HDPE) ribbon of rectangular cross-section was used as the core and the surrounding air acted as the cladding [56]. Though the achieved loss was low (~ 1 cm$^{-1}$ over 0.1 – 3.5 THz) for 10's of mm length of ribbon, dispersion was too high (1 ps pulse broadens to 20 – 40 ps, depending on the number of cycles of oscillation). The first circular core dielectric step index fiber for THz guidance was proposed in 2006 [158]. It was just a simple plastic (PE) wire of sub-wavelength diameter (200 μm) that acted as the fiber core with the surrounding air acting as the cladding. This structure supports a single mode ($HE_{11}$) at frequencies of up to 0.3 THz. Due to the small value of the core diameter, the mode spreads considerably into the surrounding air, which essentially reduces its dielectric loss (~ 0.01 cm$^{-1}$). However, the mode confinement is poor which limits its propagation length, and the bending loss is also quite high. In such step index structures, the dispersion management is difficult as it depends primarily on *material* dispersion. In this context, microstructured fibers [cf. Fig. 8(b)] are quite promising, since light guiding is not only controlled by index contrast but can be easily manipulated through multi-parameters of the wave guiding geometry, and the total dispersion becomes a strong function of *waveguide* dispersion [132]-[134]. In 2002, *Han et al.* experimentally demonstrated a HDPE based solid-core microstructured fiber [137], whose cladding was formed by arranging air holes on HDPE matrix in a periodic triangular fashion. By tuning the structural parameters, both dispersion as well as mode confinement could be tuned appropriately. As a result, the structure offers single mode guidance over 0.1 – 3 THz with low loss and relatively low dispersion (0.5 cm$^{-1}$ and 0.3 ps/THz.cm above 0.6 THz). A Teflon based solid-core microstructured fiber was fabricated and first demonstrated for THz guidance (over 0.1 ~ 1.3 THz) in 2004 [159]. As the Teflon is a very cost effective and flexible material, it can be drawn into a longer fiber length as compared to other polymers. Moreover, the refractive index of Teflon is also low (~ 1.4 in the THz regime [135], [157]), which provides lower index contrast for air-Teflon structures and hence, the scattering loss is reduced too (see the Ref. [160], scattering loss scales as square of the index contrast). The proposed fiber in [159] shows attenuation < 0.12 cm$^{-1}$ for the targeted THz range. Recently (in 2009), COC (trade name is Topas) material based large mode area (~ 1.86 mm$^2$) and small mode area (~ 0.11 mm$^2$) solid-core microstructured fibers have been fabricated for THz guidance [161]. Both of the proposed fibers possess low loss (~ 0.09 cm$^{-1}$) over 0.35 ~ 0.65 THz, and quite low dispersion (< 1 ps/THz.cm) over 0.5 ~ 1.5 THz. Moreover, the high confinement of fiber modes makes them less bend sensitive. In search of further loss reduction, researchers proposed that the insertion of a loss-less low-index air gap (size must be less than the decay length of the operating wavelength) inside the solid core can reduce the material absorption loss significantly [156]. In such a configuration, the discontinuity of the normal component of the electric field at the dielectric interface enhances the field significantly (proportional to the square of the ratio of high and low refractive index) inside the low-index region (here air gap), which results in a lower material loss. The achieved loss in air-gap silica fiber is quite small (~ 0.12 cm$^{-1}$ below 0.6 THz), however it increases rapidly at higher frequencies. The study also revealed that, with the best possible design, only 26% power can be coupled into the air hole region. In 2008, two research groups, *Hassani et al.* [138], [162] and *Atakaramians et al.* [163] theoretically proposed a *porous-core fiber* (PoCF) structure [cf. Fig. 8(c)] for efficient THz guiding, where instead of one air-hole, a two-dimensional periodic array of air holes is included in the core region. In both cases, the PoCF structures are optimized to get maximum power fraction in the air holes. One group investigated the Teflon-based PoCF, where loss (absorption + bending) was less than 0.02 cm$^{-1}$ at 1 THz with bend-radii as tight as 3 cm [138]. The second group proposed PMMA based PoCF with maximum achievable porosity of ~ 74%. Both power fraction and losses are studied in detail in [163]. Recently, a plastic (Teflon)-based *porous-core* with a *porous-cladding* fiber has been proposed [164], where air holes are arranged in hexagonal periodic fashion in both the core and the cladding regions. To support the IG-mechanism, the air-filling fraction of the core was chosen lower than that of cladding. The variation of mode-size areas (spot size, $A_{eff}$, $A_{SMI}$) and effective index ($n_e$) of the fundamental mode ($H^x_{11}$) with the outer pitch ($\Lambda_o$) is shown in Fig. 9(a). The air hole diameters and their separation are denoted as $d_i$ and $\Lambda_i$ for core region, and $d_o$ and $\Lambda_o$ for cladding region, respectively. Figure 9(b) shows the variation of power confinement in the core air-holes with $\Lambda_o$. It takes maximum value for $\Lambda_o$ ~ 0.35 mm. The maximum confinement > 20% is achievable in the *core air-holes* for $d_i/\Lambda_i = 0.6$ and $d_o/\Lambda_o = 0.8$, however, the total power in combined air-holes (core + clad) reaches up to 60% with $d_i/\Lambda_i = 0.85$ and $d_o/\Lambda_o = 0.95$, which will reduce the modal loss of this waveguide by 60% of the material loss. The numerical analysis of such a PoCF is relatively easy, however their fabrication is not. In 2009, PE-based PoCF was fabricated by *Dupuis et al.* [165] and PMMA-based PoCF was fabricated and characterized by *Atakaramians et al.* [166]. In the former work, the authors fabricated a PE rod with one layer of circular air holes [cf. Fig. 10(a)]. The achieved porosity was ~ 40% and the measured loss was ~ 0.01 cm$^{-1}$ in the vicinity of 0.3 THz. In the latter work, the authors fabricated and characterized two PMMA-based porous-core structures, one had a rectangular geometry [cf. Fig. 10(b)] and the other had a spider-web structure [cf. Fig. 10(c)]. With these geometries, not only the porosity increased (for rectangular: 65% and for spider-web: 57%), but a strong birefringence (~ 0.012 at 0.65 THz) also got introduced. Several other proposals have been made in the literature on *solid*-and *porous*-core IG-fibers, such as PoCF with ultra-high porosity (~ 86%) [167], spider-web PoCF with ultra-low loss and low-dispersion [168], randomly porous fiber with improved properties [169], suspended-solid and porous core fiber [170], PoCF with elliptical holes with high birefringence (~ 0.0445) [171], rotated porous-core microstructured fiber [172], graded index PoCF [173] etc. We have summarized the reported IG-fibers for THz transmission in Table II.

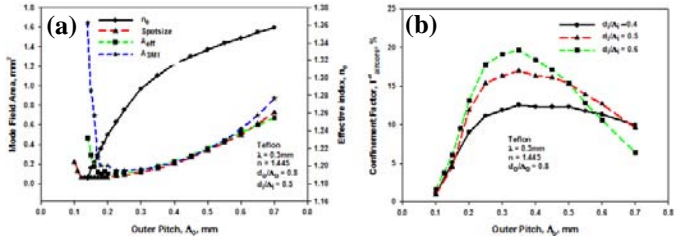

Fig. 9. The variation of (a) mode-size areas and mode effective index and (b) confinement factor in core air-holes of fundamental mode with outer pitch ($\Lambda_o$) for $d_o/\Lambda_o = 0.8$ and $\lambda = 0.3$ mm. After, Ref. [164].

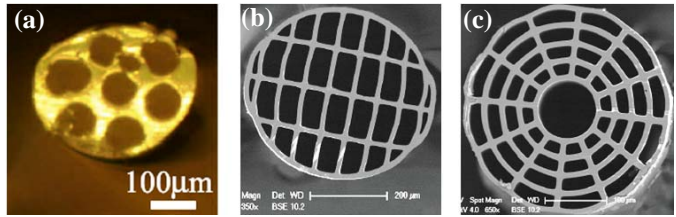

Fig. 10. Cross-sectional image of fabricated porous-core fibers. (a) PE-based circular air hole structure. Reprinted from Optics Exp., 2009, Ref. [165]. PMMA-based (b) rectangular air hole structure and (c) spider-web air hole structure. Reprinted from Optics Exp., 2009, Ref. [166].

TABLE II
THz TRANSMISSION THROUGH INDEX-GUIDED FIBERS

| Structure | Material | Core diameter | Loss (cm$^{-1}$) | Dispersion | Ref. (year) |
|---|---|---|---|---|---|
| Bare plastic ribbon | HDPE | 2 cm × 150 μm | ~ 1 (0.1 – 3.5 THz) | 1 ps → 20 - 40 ps | [56] (2000) |
| Plastic wire | PE | 200 μm | ~ 0.01 at 0.3 THz | ----- | [158] (2006) |
| Solid-core microstructured clad | HDPE + air holes | 500 μm | ~ 0.5 (0.1 – 3 THz) | -0.3 ps/THz.cm (> 0.6 THz) | [137] (2002) |
| -do- | Teflon + air holes | 1 mm | < 0.12 (0.1 – 1.3 THz) | ----- | [159] (2004) |
| -do- | COC + air holes | 870 μm, 420 μm | ~ 0.09 (0.35 – 0.65 THz) | < 1 ps/THz.cm (0.5 – 1.5 THz) | [161] (2009) |
| Solid-core with an air gap | silica | ~ 182 μm | ~ 0.12 (< 0.6 THz) | ----- | [156] (2006) |
| Porous core fiber | Teflon + circular air holes | ~ 340 μm | < 0.02 at 1 THz | ----- | [138] (2008) |
| -do- | PMMA + -do- | 400 μm | ~ 0.01 at 0.2 THz | ----- | [163] (2008) |
| -do- | PE + -do- | 350 μm | ~ 0.01 at 0.3 THz | ----- | [165] (2009) |
| -do- | PE + -do- | 445 μm, 695 μm | < 0.02 (0.1 – 0.5 THz) | < 1 ps/THz.cm (0.1 – 0.5 THz) | [167] (2010) |
| -do- | PMMA+ rectangular holes | 350 μm | ----- | $N_{eff}$ < 1 (0.35 – 0.8 THz) | [166] (2009) |
| -do- | COC + spider web | 600 μm | < 0.08 (0.2 – 0.35 THz) | -1.3 to -0.5 ps/m/μm (0.2 – 0.35 THz) | [168] (2011) |
| -do- | COC + elliptical air holes | 280 μm | < 0.05 (0.8 – 1.2 THz) | ----- | [171] (2013) |
| Randomly porous core | Zeonor + circular air holes | 390 μm | ~ 0.06 with 50% porosity | -1 to 0.1 ps/m/μm (0.8 – 1.4 THz) | [169] (2010) |
| Porous core and clad | Teflon + -do- | ~ 350 μm | ~ 0.12 (1 THz) (bend) | ----- | [164] 2012 |
| Rotated porous core | COC (Topas) | ~ 300 μm | ~ 0.066 at 1 THz | 1.06 ± 0.12 ps/THz.cm (0.5 – 1.08 THz) | [172] (2015) |
| Graded porous core | PE | 1.35 mm | 0.025 – 0.15 (.3 – 1.5 THz) | ~ 1 ps/THz.cm (0.2 – 1.5 THz) | [173] (2015) |
| Suspended solid and porous core | PE + air holes | 150 μm and 900 μm | ~ 0.02 (0.28 – 0.48 THz) | ----- | [170] (2011) |

Apart from IG-fiber structures, several other dielectric-based fiber geometries have been proposed and demonstrated. Unlike IG-fibers, light can be guided in the *lower* refractive index region of such fibers. Since the core is composed of a lower refractive index than the surrounding cladding, it allows the use of *air* (the best material for THz guidance) as the core material to further reduce the loss as well as dispersion of the transmitted THz wave. The fundamental guiding mechanisms in such fibers are, *PBG-guidance, anti-resonance reflective (ARR)-guidance* and sometimes simply *reflective guidance*.

In the case of a *PBG-guided fiber*, the dielectric cladding consists of periodic refractive index variations. As a result of that, a certain range of frequencies does not propagate across it owing to the formation of a photonic band-gap [174]. Introduction of a suitable *defect* region in that otherwise periodic structure would allow the band-gap mode to transmit through it. The PBG-fibers can be classified in two broad categories depending on the type of periodicity in the cladding structure. One is *Bragg-fiber* [175], [176] where the cladding is formed by a series of concentric alternate high and low refractive index layers [cf. Fig. 11(a)]. The other type is *holey* fiber [133], [177] [cf. Fig. 11(b)], where the cladding is formed by high and low refractive index inclusions in the form of various lattice patterns, such as, triangular, hexagonal, honeycomb etc. The transmission loss in such structures is decided primarily by leakage and material absorption loss. Though the dispersion is negligible around the centre of band-gap, it increases rapidly at the band-edges. This technology is well established in the optical domain [177], however, it is still an emerging one for THz applications.

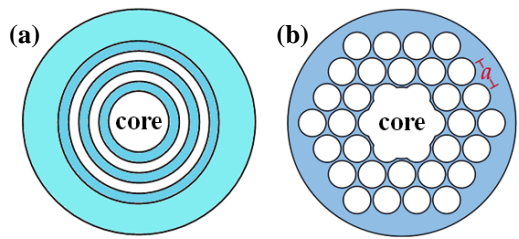

Fig. 11. Schematic cross-sectional diagram of a (a) Bragg-fiber with 1-D periodicity in RI, (b) Hollow core PBG-holey fiber with 2-D periodicity in RI in the cladding. Central core is made of air or any other lower index material.

The first proposal of air-core PBG-guided fiber for THz transmission was reported by *Skorobogatiy et al.* in 2007 in the form of an all-polymer Bragg-fiber [178]. The periodic cladding is formed by repeating alternate concentric layers (total 31 layers) of ferroelectric polyvinylidene fluoride (PVDF) and polycarbonate (PC) polymer. Over the frequency range of 0.6 to 2 THz, the PVDF acts as reflecting metal surface and over the range 2 to 2.6 THz, the refractive index contrast of PVDF/PC increases sufficiently to provide strong band-gap. As a combined effect, the structure shows low-loss transmission for THz waves over a wide window (< 0.02 cm$^{-1}$ over 1 ~ 3 THz). Another theoretical result on a cobweb-structured Bragg-fiber [cf. Fig. 12(a)] was reported in 2007 [179], where the cladding consists of alternate layers of air and a polymer material (HDPE) with supporting polymer bridges. The structure supports lowest loss mode, TE$_{01}$, which is non-Gaussian, along with few lossy higher order modes. The achievable loss for fundamental mode is quite low, < 1.9×10$^{-5}$ cm$^{-1}$ over 0.3 to 4.3 THz. An air-Teflon multilayer Bragg-fiber with supporting circular bridges [cf. Fig. 12(b)] was proposed in 2008 [138], where very low loss (< 0.02 cm$^{-1}$ around 1 THz) can be achieved with just 3 layers of material. In all these studies no dispersion value was reported. First experimental demonstration of a Bragg-fiber for THz wavelength was presented by *Ponseca et al.* in 2008 [180].

They designed and fabricated hollow core Bragg-fibers with a circular ring of air hole [cf. Fig. 12(c)]. These rings effectively act as high and lower index layers for the Bragg reflections from each interface to occur in phase [176]. They fabricated two similar fibers, where one fiber was 10% smaller in cross-section than the other. By using THz-TDS setup, the propagation of THz wave was studied, and the achieved losses were ~ 1.3 cm$^{-1}$ (over 0.8 – 1.4 THz) for one fiber and ~ 1.1 cm$^{-1}$ (over 1 – 1.6 THz) for the other. Recently, two types of air-core Bragg-fiber were designed and fabricated [181], one is air-Teflon based [similar to Fig. 12(b)] and the other is a PE and TiO$_2$-doped PE based multi-layer structure. The authors have made experimental and numerical studies on both fiber structures and shown that both of them possesses quite low loss (< 0.3 cm$^{-1}$) over 0.1 – 2 THz and low dispersion (< 1 ps/THz.cm) around center of band-gaps.

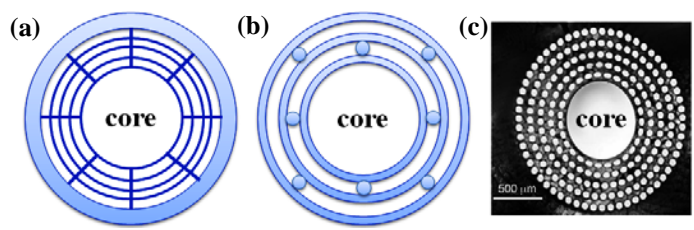

Fig. 12. Schematic cross-sectional diagram of hollow core Bragg-fibers with (a) Cob-web structure and (b) circular bridges. White regions are air and shaded regions are plastics. (c) Microscopic image of fabricated Bragg-fiber with circular ring of air holes. Reprinted from Opt. Lett., 2008, Ref. [180].

In parallel to the Bragg-fiber, several proposals on *PBG-holey* fibers have been reported in the literature. A numerical analysis of power transmission and dispersion was carried out by *Geng et al.* in 2008 [182], where they studied a HDPE based structure whose cladding was formed by 5 rings of circular air holes in a hexagonal pattern and the core was formed with a larger air hole [similar to Fig. 11(b)]. The proposed structure, with a core diameter of ~ 600 μm and an air-filling fraction of 0.9, achieved over the 1.55 ~ 1.85 THz frequency range, total loss < 0.025 cm$^{-1}$ and dispersion < ± 1 ps/THz.cm. A similar theoretical analysis has been carried out for Teflon as well as HDPE based hollow-core fiber composed of 3 cladding rings [183]. However, the chosen core diameter (~ 2.7 mm) was quite large as compared to the previous one. Due to this larger core, the structure supported unwanted higher order modes also. The Teflon-based fiber yields wider band-gap (~ 250 GHz centered at 1 THz) than the HDPE-based fiber (~ 190 GHz centered at 0.9 THz), and over this band-gap transmission loss and dispersion were < 0.01 cm$^{-1}$ and < 2 ps/km.nm, respectively. Recently, we have proposed a design of a Teflon-based air core PBG-holey fiber [similar to the Fig. 11(b)] for broad-band (> 200 GHz) low-loss THz transmission [184]. Main motivation was to reduce the transmission loss, dispersion, as well as the overall fiber dimension. As the air-Teflon structure provides relatively low refractive index contrast (1:1.44), the band-gap appears only for suitable non-zero values of longitudinal wave vector ($k_z$, along fiber length) in such simple hexagonal cladding geometry (cf. Fig. 13) [174], [184]. For a core diameter of ~ 840 μm with only 3 cladding rings, the transmission loss was < 0.04 cm$^{-1}$ [cf. Fig. 14(a)] and total dispersion was < 10 ps/km.nm over 1.65 -1.95 THz (300 GHz) [cf. Fig. 14(b)].

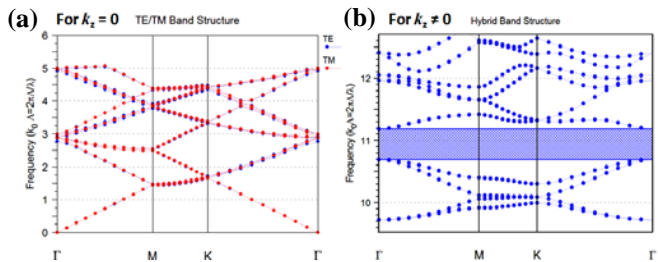

Fig. 13. Plot of normalized frequency (Λ/λ) vs in-plane wave-vector (k) for (a) $k_z = 0$ and (b) $k_z \neq 0$. The Γ, M and K are the coordinates of k.

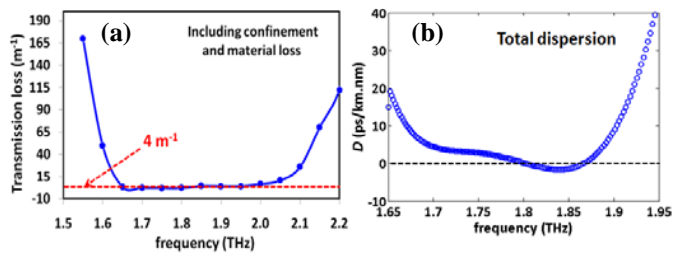

Fig. 14. Air-Teflon PBG-holey fiber. (a) Transmission loss and (b) total dispersion vs frequency.

Though detailed theoretical analyses have been made on hollow-core PBG-holey fibers, their experimental realization was quite difficult. Recently, *Bao et al.* designed and fabricated a *porous-core* PBG-holey fiber [185]. They employed a honeycomb cladding lattice and introduced a *porous-core* with a similar hole size as in cladding [cf. Fig. 15(a)]. With a high air-filling fraction, the porous core acts similar to a hollow core, while simultaneously, it offers easy fabrication, higher precision and better reproducibility of the structure. THz-TDS was used to investigate this fiber's properties. It was found that, with average diameter of 800 μm, the fiber possess quite low loss (< 0.35 cm$^{-1}$) over the 0.75 – 1.05 THz band. Very recently a comprehensive study was carried out by *Fan et al.* [186] on a similar honeycomb cladding based porous-core PBG fiber. They modeled such complex structure in a simple equivalent step index fiber form and developed a semi-analytical method to solve the modal properties quite accurately. Another theoretical study was carried out by *Liang et al.* [187] on a similar porous-core PBG structure, except that the cladding geometry was hexagonal and the hole size and pitch were different for the core and cladding regions. They observed for the fiber with the core diameter of ~ 850 μm and 3 cladding rings, the transmission loss was < 0.023 cm$^{-1}$ over 1.025 ~ 1.2 THz and dispersion was < ± 2.5 ps/THz.cm over 0.98 – 1.15 THz. Another very interesting *hollow-core* PBG-holey fiber with *hyperuniform disordered* cladding geometry [cf. Fig. 15(b)] was proposed recently by *Laurin et al.* [188]. In a hyperuniform material structural uniformity can be varied from 0 (disordered) to 1

(crystal). The authors have already reported its numerical analysis, and fabrication is apparently under way.

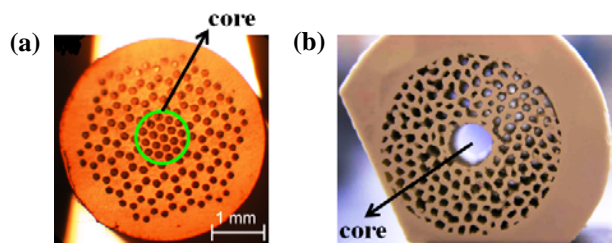

Fig. 15. Micrograph of the fabricated PBG-fiber. (a) Porous-core honeycomb structure. Reprinted from Optics Exp., 2012, Ref. [185]. (b) Hollow-core hyperuniform disordered structure. Reprinted from Proc. IEEE IRMMW-THz, 2014, Ref. [188].

Apart from PBG-guidance, researchers have explored few other guiding mechanisms to confine the THz mode within the hollow-core. These include *anti-resonance-reflective* (ARR) guidance and metallic type reflection in single clad structure. Under the ARR effect, both the core and the microstructured cladding support propagating modes, but they do not couple to each other owing to the anti-resonance effect. Some examples are, hollow-core with a periodic microstructured cladding [189], a hollow-core kagome structure [190], and a hollow-core with tube-lattice cladding [191]. Similar to a metallic WG, single clad plastic fiber for THz guidance is also proposed by *Hidaka et al.* [58], [192], where they have used a PVDF tube to guide the THz mode inside the air core. Recently, a hollow-core fiber with two and four embedded metal (Indium) wires has been fabricated for low-loss THz guidance [193]. Table III summarizes the reported PBG and ARR-guided dielectric fibers for THz wave.

TABLE III
THz TRANSMISSION THROUGH PBG & ARR-GUIDED FIBERS

| Structure | Material | Core dia. | Loss (cm$^{-1}$) | Dispersion | Ref. (year) |
|---|---|---|---|---|---|
| All polymer Bragg fiber | PVDF + PC | 1 mm | < 0.02 (1 - 3 THz) | ------ | [178] (2007) |
| Spider-web Bragg fiber | HDPE + air | 16, 20, 30 mm | < 1.9e$^{-5}$ (0.3 – 4.3 THz) | ------ | [179] (2007) |
| Bragg with circular bridges | Teflon + air | 2 mm | < 0.02 at ~ 1 THz | ------ | [138] (2008) |
| Bragg with circular ring of holes | PMMA + air | 670 μm | ~ 1.1 (1-1.6 THz) | ------ | [180] (2008) |
| Bragg fiber multilayer | Teflon + air & PE + TiO$_2$ doped PE | 6.73 & 6.63 mm | < 0.3 (0.1 – 2 THz) | < 1 ps/THz.cm inside PBG | [181] (2011) |
| PBG- holey fiber | HDPE + air | 584 μm | < 0.025 (1.55 – 1.84 THz) | < ± 1 ps/THz.cm (1.63 – 1.8 THz) | [182] (2008) |
| -do- | Teflon + air & HDPE + air | ~ 2.7 mm | < 0.01 (0.8 ~ 1.1 THz) | < 2 ps/km.nm (0.9 – 1.1 THz) | [183] (2009) |
| -do- | Teflon + air | 840 μm | < 0.04 (1.65 – 1.95 THz) | < ± 2 ps/km.nm (1.7 – 1.9 THz) | [184] (2014) |
| PBG-porous-core honeycomb | COC (Topas) + air | 800 μm | < 0.35 (0.75 – 1.05 THz) | Pulse broadened by 20 times | [185] (2012) |
| Porous-core hexagonal PBG | -do- | ~ 850 μm | < 0.023 (1.025 – 1.2 THz) | < ± 2.5 ps/THz.cm (0.98- 1.15 THz) | [187] (2013) |
| PBG-hyperuniform clad | ------ | 6 mm | ------ | ------ | [188] (2014) |
| ARR-microstructured fiber | Teflon + air | ~ 5.5 mm | 0.01 at ~ 0.8 THz | ------ | [189] (2008) |
| ARR-Kagome lattice | PMMA + air | 1.6 & 2.2 mm | 1 (0.75 – 1 THz) & 0.6 (0.65 – 1 THz) | < ± 10 ps/THz.cm (0.8/0.65 – 1.05 THz) | [190] (2011) |
| ARR- tube lattice | Zeonex + air | 3.84 mm | < 0.003 (0.36-1.09 THz) | $N_{eff}$ (0.99 – 1) for 0.4 – 1.2 THz | [191] (2015) |
| Single clad pipe | PVDF + air | 8 mm | ~ 0.015 (1 – 2 THz) | ------ | [192] (2005) |
| Hollow core with embedded metal wire | Zeonex + Indium | ~ 2 mm | ~ 0.3 & 0.6 (0.55 – 1 THz) | ~ 5 ps/THz.cm (0.65 – 1 THz) | [193] (2013) |

## IV. CONCLUSION

The use of optical fibers for THz applications has attracted considerable attention in recent years. In this article, we have reviewed the progress and current status of optical fiber based techniques for THz generation and transmission. The first part of this review focused on THz sources. After a brief discussion on various types of THz sources, we discussed how specialty optical fibers can be designed for THz generation. Most of the fiber-based THz generation schemes are based on frequency conversion via nonlinear effect of FWM. The conversion efficiency is still quite low but is likely to improve with further research.

The second part of this review focused on the guiding of THz waves for their transmission over relatively long (compared to the THz wavelength). After discussing various wave guiding schemes, we discuss many fiber designs suitable for THz transmission. Solid-core fibers are relatively easy to fabricate and provide widest band-width for single mode operation. However, they are limited by absorption losses of the material used to make them. To minimize that loss, porous-core fibers have been proposed. As we discussed in detail, porous-core fibers show excellent properties in the THz range both in terms of their dispersion and losses. However, as a large portion of field lies in air, they are very sensitive to waveguide perturbations. Hollow-core fibers provide the lowest loss and dispersion over a specific spectral range. The main drawbacks are their limited band-width, difficulty of fabrication and a larger size. Recently researchers are exploiting interesting routes, such as using metal-dielectric hybrid structures and metamaterials to further improve the THz transmission. The field of THz radiation is likely to see considerable progress in the coming years.

**Ajanta Barh** is Ph.D. student at IIT Delhi, obtained B.Sc.' 2008 from Calcutta University and M.Sc.'2010 all in Physics from IIT Delhi. Her research interests include Specialty fibers for mid-IR and THz wavelengths, Si-photonics, Non-linear optics, guided wave components. She is student member of OSA and OSI; authored/co-authored of more than 30 articles; Received Best Student Paper award in IEEE-CODEC 2012, OSA-PHOTONICS 2014, SPIE-ICOP 2015 conferences.

**Bishnu P. Pal** is Professor of Physics at Mahindra École Centrale Hyderabad India since July 1st 2014 after retirement as Professor of Physics from IIT Delhi; Ph.D.'1975 from IIT Delhi; Fellow of OSA and SPIE; Senior Member IEEE; Honorary Foreign Member Royal Norwegian Society for Science and Arts; Member OSA Board of Directors (2009-11); Distinguished Lecturer IEEE Photonics Society (2005-07).

**Govind P. Agrawal** is Wyant Professor of Optics at the Institute of Optics, University of Rochester NY where he joined in 1989. He received his Ph.D.'74 from IIT Delhi; authored/coauthored over 400 research papers, and eight books, recipient of the IPS Quantum Electronics Award'12, Fellow of OSA, IEEE, and member OSA Board (2009-10).

**R. K. Varshney** received Ph.D. degree from the Indian Institute of Technology Delhi in 1987, where he is currently Professor of Physics. He was awarded Marie Curie Fellowship and Fullbright Fellowship. He is an author/coauthor of more than 125 research papers, and two books. He has keen interest in the fields of Fiber and Integrated Optics, Specialty fibers, Fiber Optic Sensors, Non-linear Optics.

**B. M. Azizur Rahman** is Professor City University, London. He received BS'76 and MS'79 from Bangladesh University of Engg. and Tech., Dhaka, Bangladesh and Ph.D.'82 from University College London. He is Fellow of the OSA, SPIE and a Senior Member of IEEE, has published nearly 500 journal and conference papers and specializes in the use of rigorous and full-vectorial numerical modelling of a wide range of photonic devices.